\documentstyle[prl,eqsecnum,epsf,aps]{revtex}
\begin{document}
\tighten
\draft
\twocolumn %take this out before submission
\title{A New Measurement of the Partial
$\mathrm{0^{+}\rightarrow0^{+}}$\ 
Half Life of $\mathrm{^{10}C}$\ with GAMMASPHERE.}
\author{B.K.\ Fujikawa,
S.J.\ Asztalos, R.M.\ Clark, M.-A.\ Deleplanque-Stephens, P.\ Fallon,
S.J.\ Freedman, I-Y.\ Lee, L.J.\ Lising, A.O.\ Macchiavelli,
R.W.\ MacLeod, J.C.\ Reich,
M.A.\ Rowe, S.-Q.\ Shang, F.S.\ Stephens, and E.G.\ Wasserman}
\address{University of California and the
Lawrence Berkeley National Laboratory, Berkeley, CA 94720}
\author{J.P.\ Greene}
\address{Argonne National Laboratory, Argonne, IL 60439}
\date{\today}
\maketitle
\begin{abstract}
We report on a new measurement of the strength of the 
superallowed \mbox{$\mathrm{0^{+}\rightarrow0^{+}}$} transition
in the $\beta$-decay of $\mathrm{^{10}C}$:
\mbox{$\mathrm{^{10}C(0^{+},g.s.)
\rightarrow^{10}B(0^{+},1.74MeV)+e^{+}+\nu}$}.
The experiment was done at the LBNL 88-inch cyclotron
using forty seven GAMMASPHERE germanium detectors.
Precise knowledge of this branching ratio is necessary to compute
the superallowed Fermi $\mathrm{{\mathit{f}}t}$, which gives
the weak vector coupling constant and the u to d element of the
Cabibbo-Kobayashi-Maskawa quark mixing matrix.
\end{abstract}

The most precise value of the u to d element of the
Cabibbo-Kobayashi-Maskawa (CKM) quark mixing matrix is obtained from
measurements of superallowed $\mathrm{0^{+}\rightarrow0^{+}}$\ Fermi
$\beta$-decays in nuclear systems.
Specifically, these decay rates determine the nucleon weak vector
coupling constant $\mathrm{G_{V}}$\ giving $\mathrm{V_{ud}}$:
\mbox{
$\mathrm{G_{V}^{2}=G_{F}^{2}\left|V_{ud}\right|^{2}(1+\Delta_{R})}$}
where $\mathrm{G_{F}}$\ is the Fermi coupling constant obtained from
the muon lifetime and $\mathrm{\Delta_{R}}$\ is a nucleus independent
(``inner'') radiative correction.
The conserved vector current (CVC) hypothesis implies that
superallowed $\mathrm{{\mathit{f}}t}$-values within isospin-1
multiplets are related to $\mathrm{G_{V}}$\ by:
\begin{equation}
\mathrm{{\mathit{f}}(1+\delta_{R})(1-\delta_{C})t
\equiv{\mathcal{F}}t=
\frac{K}{G_{V}^{2}\left|M_{V}\right|^{2}}}
\label{eq:ft}
\end{equation}
where \mbox{$\mathrm{\left|M_{V}\right|^{2}=2}$} is the vector
matrix element,
$\mathit{f}$\ is the familiar Fermi statistical rate function,
$\mathrm{\delta_{R}}$\ is the nucleus dependent (``outer'') radiative
correction, $\mathrm{\delta_{C}}$\ is the charge dependent
correction to \mbox{$\mathrm{\left|M_{V}\right|^{2}=2}$} due to
isospin symmetry breaking, and K is the usual $\beta$-decay constant.
The corrected $\mathrm{{\mathcal{F}}t}$ include nuclear and
radiative effects.
Precise determination of $\mathrm{G_{V}}$\ requires precision
measurements of the partial
\mbox{$\mathrm{0^{+}\rightarrow0^{+}}$} half-life, the
$\beta$\ endpoint energy, and reliable theoretical calculations of
$\mathrm{\delta_{R}}$\ and $\mathrm{\delta_{C}}$.
Reference \cite{EHagberg96} summarized the status of measurements and
calculations for $\mathrm{^{10}C}$, $\mathrm{^{14}O}$,
$\mathrm{^{26}Al^{m}}$, $\mathrm{^{34}Cl}$, $\mathrm{^{38}K^{m}}$,
$\mathrm{^{42}Sc}$, $\mathrm{^{46}V}$, $\mathrm{^{50}Mn}$,
and $\mathrm{^{54}Co}$.
The constancy of $\mathrm{{\mathcal{F}}t}$ for these nine precisely
measured superallowed decays supports the CVC hypothesis.
This review suggests
\mbox{$\mathrm{\left|V_{ud}\right|=0.9740\pm0.0005}$}.
Together with the two other elements in the first row of the CKM
matrix taken from ref.\ \cite{PDG96}, this tests the unitary of the
CKM matrix.
The result,
\mbox{$\mathrm{\left|V_{ud}\right|^{2}+\left|V_{us}\right|^{2}+
\left|V_{ub}\right|^{2}\equiv\left|V\right|^{2}=0.9972\pm0.0013}$},
is more than two standard deviations from the unitary constraint.

\begin{figure}
\begin{center}
\epsfxsize=0.44\textwidth 
\epsffile{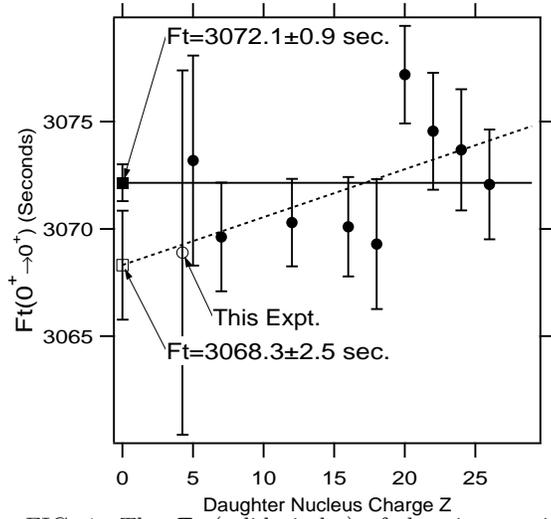}
\caption{The $\mathrm{{\mathcal{F}}t}$\ (solid circles) of the nine
precisely measured superallowed decays ($\mathrm{^{10}C}$,
$\mathrm{^{14}O}$, $\mathrm{^{26}Al^{m}}$, $\mathrm{^{34}Cl}$,
$\mathrm{^{38}K^{m}}$, $\mathrm{^{42}Sc}$, $\mathrm{^{46}V}$,
$\mathrm{^{50}Mn}$, and $\mathrm{^{54}Co}$) plotted as a function
of the daughter nucleus charge Z.\ \ 
The solid line is the weighted average.
The dashed line is the result of a linear fit and the open square
is the extrapolation of this fit to zero charge.
The open circle is the
$\mathrm{^{10}C}$\ $\mathrm{{\mathcal{F}}t}$\ using the
superallowed branching ratio from this measurement.}
\label{fig:FtValues}
\end{center}
\end{figure}

A violation of CKM unitary requires the Standard Model to be
extended.
A more mundane explanation is unaccounted systematic uncertainties in
the difficult theoretical calculations needed to extract
$\mathrm{V_{ud}}$.
The calculation of the isospin symmetry breaking correction
$\mathrm{\delta_{C}}$\ is regarded as the most problematic.
Figure \ref{fig:FtValues} shows the most precisely measured
$\mathrm{{\mathcal{F}}t}$\ as a function of daughter nucleus charge
Z.\ \ 
Possible unaccounted Z-dependent corrections motivated
extrapolations to zero charge using second and third order polynomials
fits to $\mathrm{\delta_{C}}$-corrected\cite{GSavard95} or
uncorrected\cite{DHWilkinson93} $\mathrm{{\mathit{f}}t}$ values.
The agreement of the extrapolated values with unitary suggests that
incomplete isospin corrections might explain the discrepancy.
The $\mathrm{{\mathit{f}}t}$\ for the superallowed Fermi
$\beta$\ decay of $\mathrm{^{10}C}$\ is of particular interest:
$\mathrm{^{10}C}$\ has the lowest nuclear charge of a superallowed
Fermi decay.
Moreover, all existing calculations agree that
$\mathrm{\delta_{C}}$\ for $\mathrm{^{10}C}$ is small enough to be
neglected.

The necessary experimental input are the total half-life, the
branching fraction for
\mbox{$\mathrm{^{10}C(0^{+},g.s.)\rightarrow}$}
\mbox{$\mathrm{^{10}B(0^{+},1.74MeV)+e^{+}+\nu}$},
and the superallowed endpoint energy.
The half-life ($19.290\pm0.012$\ seconds\cite{PHBarker90}) and the
recently revised endpoint energy
($885.86\pm0.12$\ keV\cite{SCBarker89,PHBarker98})
are known to high precision; the limiting experimental
input is the \mbox{$\mathrm{0^{+}\rightarrow0^{+}}$} branching ratio.
Figure \ref{fig:B10Level} shows the $\mathrm{^{10}B}$\ and
$\mathrm{^{10}C}$\ levels important for this measurement.
The $\beta$\ decay of $\mathrm{^{10}C}$\ goes to the
\mbox{$\mathrm{^{10}B(0^{+},1.740 MeV)}$} or the
\mbox{$\mathrm{^{10}B(1^{+},0.718 MeV)}$} state.
The allowed decay to the \mbox{$\mathrm{^{10}B(1^{+},2.154 MeV)}$}
level is known to be small experimentally ($<8\times10^{-6}$) as
expected from the meager available energy.
The forbidden $\beta$\ decay to the $\mathrm{^{10}B}$\ ground state is
suppressed by about $10^{-10}$.
The decay to the \mbox{$\mathrm{^{10}B(0^{+},1.740 MeV)}$} state
is followed with $\gamma$-rays at 1022 keV and 718 keV.
The direct ground state decay of the
\mbox{$\mathrm{^{10}B(0^{+},1.740 MeV)}$} level is magnetic octupole,
with an estimated branch below $10^{-12}$\cite{MAKroupa91}.
The decay to the \mbox{$\mathrm{^{10}B(1^{+},0.718 MeV)}$} state is
followed by a single 718 keV $\gamma$-ray.
Therefore the \mbox{$\mathrm{0^{+}\rightarrow0^{+}}$} branching ratio
is the same as the $\gamma$-ray intensity ratio:
\begin{equation}
\mathrm{
b=\frac{I_{\gamma}(1022keV)}{I_{\gamma}(718keV)}
=\frac{Y(1022keV)}{Y(718keV)}
\frac{\epsilon(718keV)}{\epsilon(1022keV)}}
\label{eq:BR}
\end{equation}
where $\mathrm{Y(\gamma)}$\ is the $\gamma$-ray yields from
$\mathrm{^{10}C}$\ $\beta$-decay and $\epsilon(\gamma)$\ full energy
$\gamma$-ray detection efficiencies.

This experiment was performed with the GAMMASPHERE\cite{IYLee90}
detector at the Lawrence Berkeley National Laboratory 88-Inch
Cyclotron.
Three measurements are required: a measurement of the $\gamma$-ray
yield ratio following $\beta$-decay, the full energy
$\gamma$-ray detection efficiency ratio, and the
$\mathrm{2\times511}$\ keV pileup background to the 1022 keV
$\gamma$-ray peak.
For the $\beta$-decay measurement, the $\mathrm{^{10}C}$\ source is
produced with the $\mathrm{^{10}B(p,n)^{10}C}$\ reaction using a
$\mathrm{325\mu g/cm^{2}}$\ thick target of 99.5\% enriched
$\mathrm{^{10}B}$\ on a $\mathrm{600\mu g/cm^{2}}$\ thick carbon
backing and a 250 nA 8 MeV proton beam.
The $\beta$\ delayed $\gamma$-rays from $\mathrm{^{10}C}$\ decay were
detected by forty-seven GAMMASPHERE germanium detectors.
The usual BGO Compton suppressors were turned off in order to avoid
possible systematic effects from ``false vetoes'' by an unrelated
$\gamma$-ray.
A 35 second beam-on/beam-off cycle with a 1 second delay was used.
We use a technique employed by a previous experiment\cite{GSavard95}
for measuring the $\gamma$-ray efficiency ratio.
The efficiency is measured in situ with the $\gamma$-rays of
interest by tagging $\gamma$\ cascades prepared by exciting
the \mbox{$\mathrm{^{10}B(1^{+},2.154 MeV)}$} state.
A reduced intensity 10 nA proton beam is used to populate this state
with \mbox{$\mathrm{^{10}B(p,p')^{10}B^{*}}$}.
The
\mbox{$\mathrm{^{10}B(1^{+},2.154 MeV)\rightarrow}$}
\mbox{$\mathrm{^{10}B(0^{+},1.740 MeV)}$}
transition is tagged with the 414 keV $\gamma$-ray.
The \mbox{$\mathrm{^{10}B(0^{+},1.740 MeV)}$} state then decays to
the $\mathrm{^{10}B}$\ ground state by emitting exactly one 1022 keV
$\gamma$-ray and one 718 keV $\gamma$-ray.
The distribution of these $\gamma$-ray is isotropic because the
cascade begins with the $0^{+}$\ state.

\begin{figure}
\begin{center}
\epsfxsize=0.44\textwidth 
\epsffile{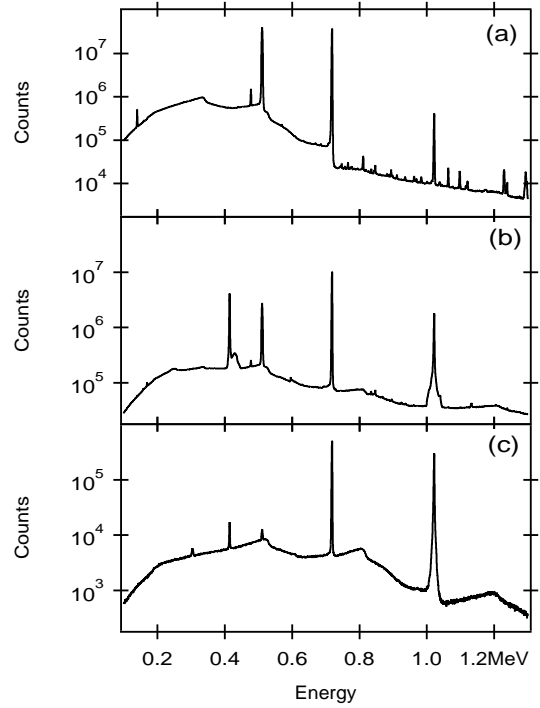}
\caption{(a) The $\beta$-delayed $\gamma$-ray energy spectrum.
(b) The prompt $\gamma$-ray energy spectrum.
(c) The prompt $\gamma$-ray energy spectrum gated by the 414 keV
$\gamma$-ray.
The peaks at 718 keV and 1022 keV in the delayed spectrum are from
the $\beta$-decay of $\mathrm{^{10}C}$.
The remaining peaks are due to positron annihilation, room background,
neutron activation, and background proton reactions, primarily:
$\mathrm{^{10}B(p,\alpha)^{7}Be}$.
The peaks at 718 keV and 1022 keV in the gated spectrum are used to
measure the relative efficiency.
The residual 414 keV $\gamma$-ray peak in (c) disappears after
corrections are made for accidentals and Compton background.}
\label{fig:EHIST}
\end{center}
\end{figure}

Figure \ref{fig:EHIST}a shows the $\beta$-delayed $\gamma$-ray energy
spectrum.
We use the following procedure to determine the $\gamma$-ray yields.
The region around the $\gamma$-ray peak is fit to a function imitating
the peak and a smooth underlying background.
The peak is modeled by a Gaussian having small exponential tails and
the background is taken as a quadratic polynomial with a resolution
smoothed step function.
The step function accounts for the discontinuity in the background
caused by scattering of $\gamma$-rays in inactive material in front
of the detector.
Peaks for background radiation are included.
The fit is performed by minimizing a $\Xi$-square
function\cite{SBaker84} using MINUIT\cite{FJames75}.
The fitting procedure is used only for determining the background;
the yields are computed by subtracting the fitted background from the
data.

Figure \ref{fig:EHIST}b shows the $\gamma$-ray spectrum from the
\mbox{$\mathrm{^{10}B(p,p')^{10}B^{*}}$} reaction.
The gating process is as follows.
A fit to the 414 keV peak is performed using the method described.
The result is used to determine the energy window, defined to be
$\pm1\sigma$\ centered about the peak.
The 414 keV peak is on a smooth background which includes Compton
scattering of higher energy $\gamma$-rays.
The effect of this Compton background is estimated by taking eleven
additional energy gates below and twelve above the 414 keV peak.
The Compton background is determined for each gate and a quadratic
polynomial interpolation is used to estimate the Compton background
under the 414 keV $\gamma$-ray peak.
The background under the 414 keV peak also includes a small double
escape peak from the 1436 keV $\gamma$-ray which is emitted in the
\mbox{$\mathrm{^{10}B(1^{+},2.154 MeV)\rightarrow}$}
\mbox{$\mathrm{^{10}B(1^{+},0.718 MeV)}$}
transition.
Since the 1436 keV $\gamma$-ray is always emitted with a 718 keV
$\gamma$-ray and never with a 1022 keV $\gamma$-ray, a small
correction is applied to the efficiency ratio.
This correction is determined from the number of counts in the single
escape peak and the ratio of double to single escapes from an
EGS4\cite{WRNelson85} Monte Carlo simulation.
The accidental $\gamma_{414}-\gamma$\ coincidences were corrected for
by subtracting counts obtained in non-coincident time gates.
The accidental gates were normalized to the coincidence gate by taking
advantage of the fact that it is impossible for two 718 keV
$\gamma$-rays to be in true coincidence.
The normalization factor is chosen such that the
$\gamma_{718}-\gamma_{718}$\ coincidences disappears in the
subtracted spectrum.

Since the 1022 keV $\gamma$-ray is emitted during the slow down of the
recoiling $\mathrm{^{10}B}$, a small correction is made to account for
the kinematical change in solid angle and the Doppler energy shift.
The overall correction is reduced because of the symmetry of
GAMMASPHERE.
The size of the correction was calculated with Monte Carlo
integration using the differential
\mbox{$\mathrm{^{10}B(p,p')^{10}B^{*}}$} cross sections from
ref.\ \cite{BAWatson69}, the lifetimes and cascade
branching ratios from ref.\ \cite{FAjzenberg-Selove88}, and the
stopping powers from ref.\ \cite{JFZiegler85}.

The number of background $\mathrm{2\times511}$\ keV pileup counts
in the 1022 keV $\gamma$-ray peak is measured using 
$\mathrm{^{19}Ne}$\ as a source of positrons.
The $\mathrm{^{19}Ne}$\ source is prepared in situ with the
$\mathrm{^{19}F(p,n)^{19}Ne}$\ reaction by bombarding a
$\mathrm{325\mu g/cm^{2}}$\ thick PbF target on a
$\mathrm{600\mu g/cm^{2}}$\ thick carbon foil backing with a
100 nA 8 MeV proton beam.
Like the $\mathrm{^{10}C}$\ decay measurement, a 35 second
bombardment and counting cycle was used.
The $\mathrm{^{19}Ne}$\ decay is similar to $\mathrm{^{10}C}$\ with a 
$17.239\pm0.014$\ sec.\ half-life and a similar $\beta$\ endpoint
energy ($1705.38\pm0.80$\ keV).
The $\mathrm{^{19}Ne}$\ is a source of 511 keV annihilation
$\gamma$-rays with no true 1022 keV $\gamma$-ray, and the entire peak
at 1022 keV is due to pileup.
In order to normalize the $\mathrm{^{19}Ne}$\ data to the
$\mathrm{^{10}C}$\ data, we use the following technique.
The GAMMASPHERE data stream contains a 1 MHz clock and the absolute
time of each trigger is known to 1 $\mu$s.
Using this information, we determine the number
$\mathrm{N\left(2\times511\right)}$\ of 511 keV $\gamma$-rays
within a 1 $\mu$s time bin that follow a 511 keV $\gamma$-ray with an
arbitrary delay for each detector.
Like the summing of two annihilation $\gamma$-rays this is a
purely random process.
Neglecting, for the moment, small dead time corrections
(which are later corrected for),
\mbox{$\mathrm{
N\left(2\times511\right)=
\left(R_{511}\cdot T\right)\left(R_{511}\cdot\tau_{bin}\right)}$}
where $\mathrm{R_{511}}$\ is the rate of 511 keV $\gamma$-rays in a
single detector, T is the counting time, and
\mbox{$\mathrm{\tau_{bin} = 1\ \mu s}$}
is the bin width.
Similarly, the number of $\mathrm{2\times511\ keV}$\ pileup counts in
the energy spectrum is given by:
\mbox{$\mathrm{
Y\left(2\times511\right)=
\left(R_{511}\cdot T\right)\left(R_{511}\cdot\tau_{pu}\right)}$}
where $\mathrm{\tau_{pu}}$\ is the pileup rejection time in the
GAMMASPHERE amplifiers.
The rate independent ratio
\begin{equation}
\mathrm
\frac{Y\left(2\times511\right)}{N\left(2\times511\right)}=
\frac{\tau_{pu}}{\tau_{bin}}
\label{eq:ratio}
\end{equation}
is used to compute the summing correction.

With the exception of the $\mathrm{2\times511\ keV}$ pileup, random
pileup does not effect the ratio in eqn.\ \ref{eq:BR}.
However, the correlated pileup of $\gamma$-rays from a single cascade
is a possible systematic effect.
Specifically, both the 718 keV and 1022 keV $\gamma$-rays can deposit
energy into the same detector, in effect removing counts from the
full energy peaks.
The effect cancels to first order in the efficiency ratio and an
EGS4\cite{WRNelson85} Monte Carlo simulation indicates that this
effect is $-0.032(3)\%$.
However, due to the smallness of the
$\mathrm{^{10}C}$\ \mbox{$\mathrm{0^{+}\rightarrow0^{+}}$} branch,
this effect is significant in the decay measurement.
Only about 1.5\% of the 718 keV $\gamma$-rays are emitted with
a 1022 keV $\gamma$-ray, but all 1022 keV $\gamma$-rays come
with a 718 keV $\gamma$-ray.
Systematically, more 1022 keV $\gamma$-rays will be removed from the
full energy peak.
The pileup correction for a single detector is estimated by measuring
coincidences between different detectors in the GAMMASPHERE array.
Neglecting,for the moment, threshold corrections, small variations in
detector sizes, and the small $\gamma-\gamma$\ angular correlation,
the pileup correction is equal to
\begin{equation}
\mathrm
f_{PU}=
\frac
{1+\frac{Y(1022\cdot x)}{Y(1022)}}
{1+\frac{Y(718\cdot x)}{Y(718)}}
\label{eq:CPU}
\end{equation}
where $\mathrm{Y(\gamma)}$\ is the total $\gamma$-ray yield and
$\mathrm{Y(\gamma\cdot x)}$\ is the $\gamma$-ray yield when there is
a coincident event in a second detector.
The correction for detector size and $\gamma-\gamma$\ angular
correlation are straight forward.
The detector size is simply scaled by $\mathrm{Y(718 keV)}$.
Since the transition
\mbox{$\mathrm{^{10}B(0^{+},1.740 MeV)\rightarrow}$}
\mbox{$\mathrm{^{10}B(1^{+},0.718 MeV)}$}
is pure M1 and the transition
\mbox{$\mathrm{^{10}B(1^{+},0.718 MeV)\rightarrow}$}
\mbox{$\mathrm{^{10}B(3^{+},g.s.)}$}
is primarily E2, the $\gamma-\gamma$\ angular correlation is equal to:
\mbox{$\mathrm{P\left(\cos\theta\right)=
1-0.0714\cdot P_{2}\left(\cos\theta\right)}$}
\cite{KSiegbahn68}.
The correction for threshold is more problematic.
GAMMASPHERE uses constant fraction discriminators (CFD) whose
thresholds are not easily described.
To avoid this problem, we enforce a software threshold at 417 keV,
well above the CFD threshold.
An EGS4\cite{WRNelson85} Monte Carlo simulation of GAMMASPHERE is
used to correct for the fraction of events below 417 keV.

The room background was measured for 17.5 hours after the run.
No $\gamma$-rays were found in the region of 718 keV.
However, a background $\gamma$-ray, with an energy of
\mbox{$\mathrm{1022.6\pm0.4\ keV}$}, was observed.
Based upon constraints on half-life, intensity, and associated
$\gamma$-rays, the only possible source is the $\beta$ decay of
$\mathrm{^{120}Sb}$.
This was probably produced through proton activation,
$\mathrm{^{120}Sn(p,n)^{120}Sb}$, of the aluminum alloy foil
lining the inside of the GAMMASPHERE scattering chamber.
In addition to a 1023 keV $\gamma$-ray, the $\beta$ decay of
$\mathrm{^{120}Sb}$\ emits a 197.3 keV $\gamma$-ray with nearly equal
intensity.
We use this 197.3 keV $\gamma$-ray to scale the background spectrum
to the $\mathrm{^{10}C}$\ decay data in order to subtract the
$\mathrm{^{120}Sb}$\ contamination.

The summary of all corrections are shown in table \ref{tab:BRCorrect}.
The strength of the $\mathrm{^{10}C}$\ superallowed
\mbox{$\mathrm{0^{+}\rightarrow0^{+}}$} branch is determined to be
\begin{equation}
\mathrm{
b=\left[1.4665\pm0.0038(stat)\pm0.0006(syst)\right]\times10^{-2}}
\label{eq:BRExperiment}
\end{equation}
where the systematic error dominated by the uncertainty in the
EGS4 threshold correction in the correlated pileup correction.
A comparison with previous experiments is shown in table
\ref{tab:BRCompare}.
This result is about one standard deviation from the results of
ref.\ \cite{GSavard95}.
Our result for b along with previous measurements of the
$\beta$\ endpoint energy and the total lifetime gives
$\mathrm{^{10}C}$\ \mbox{$\mathrm{{\mathcal{F}}t=3068.9\pm8.5}$\ sec.}
and \mbox{$\mathrm{\left|V_{ud}\right|=0.9745\pm0.0014}$},
using the usual radiative corrections and the isospin breaking
corrections $\mathrm{\delta_{C}=0.16(3)\%}$\ from
ref.\ \cite{EHagberg96}.
The unitary test is satisfied with data from this experiment,
\mbox{$\mathrm{\left|V\right|^{2}=0.9983\pm0.0029}$},
but the error is large.
The present experiment seems to favor a Z dependence correction of
ref.\ \cite{DHWilkinson93} but is statistics limited and a more
precise experiment is necessary to help resolve the issue.

\acknowledgments
This work was supported in part by the US DOE under contract numbers
DE-AC03-76SF00098 and W-31-109-ENG-38.

% now the references. delete or change fake bibitem. delete next three
%   lines and directly read in your .bbl file if you use bibtex.

\begin{figure}
\begin{center}
\epsfxsize=0.4\textwidth 
\epsffile{fig2.epsf}
\caption{The relevant energy levels of $\mathrm{^{10}B}$\ and
$\mathrm{^{10}C}$.
}
\label{fig:B10Level}
\end{center}
\end{figure}
\begin{table}
\caption{Summary of the experimental corrections made in the
measurement of the superallowed branching ratio.}
\label{tab:BRCorrect}
\begin{tabular}{ccc}
Correction & Size & Affects \\
\hline
Accidental Coincidences & $-(1.94\pm0.02)\%$ & Efficiency \\
Compton Background & $-(0.049\pm0.008)\%$ & Efficiency \\
Double Escape Peak & $-(0.020\pm0.004)\%$ & Efficiency \\
Kinematic Shift & $-(0.019\pm0.051)\%$ & Efficiency \\
$2\times511$ keV Pileup & $-(1.25\pm0.19)\%$ & $\beta$-decay \\
Correlated Pileup & $-(0.032\pm0.003)\%$ & Efficiency \\
& $+(0.44\pm0.05)\%$ & $\beta$-decay \\
$\mathrm{^{120}Sb}$\ Background & $-(0.23\pm0.11)\%$
& $\beta$-decay \\
\end{tabular}
\end{table}

\begin{table}
\caption{Comparison of $\mathrm{^{10}C}$\ superallowed
\mbox{$\mathrm{0^{+}\rightarrow0^{+}}$}
branching ratios.}
\label{tab:BRCompare}
\begin{tabular}{cc}
Branching Ratio & Reference \\
\hline
$(1.465\pm0.014)\times10^{-2}$ & \cite{DCRobinson72} \\
$(1.473\pm0.007)\times10^{-2}$ & \cite{YNagai91} \\
$(1.465\pm0.009)\times10^{-2}$ & \cite{MAKroupa91} \\
$(1.4625\pm0.0025)\times10^{-2}$ & \cite{GSavard95} \\
$(1.4665\pm0.0038)\times10^{-2}$ & This work \\
$(1.4645\pm0.0019)\times10^{-2}$ & World Average \\
\end{tabular}
\end{table}


\begin{references}
\bibitem{EHagberg96} E.\ Hagberg, \textit{et al.},
in \textit{Non-Nucleonic Degrees of Freedom Detected in Nucleus},
ed.\ T.\ Minamisono, \textit{et al.}, (World Scientific, 1996).
\bibitem{PDG96} Particle Data Group, Phys.\ Rev.\ \textbf{D54},
1 (1996).
\bibitem{GSavard95} G.\ Savard,
\textit{et al.}, Phys.\ Rev.\ Lett.\ \textbf{74}, 1521 (1995).
\bibitem{DHWilkinson93} D.H.\ Wilkinson,
Nucl.\ Inst.\ Meth.\ \textbf{A335}, 201 (1993).
\bibitem{PHBarker90} P.H.\ Barker, \textit{et al.},
Phys.\ Rev.\ \textbf{C41}, 246 (1990).
\bibitem{SCBarker89} S.C.\ Barker, \textit{et al.},
Phys.\ Rev.\ \textbf{C40}, 940 (1989).
\bibitem{PHBarker98} P.H.\ Barker, private communication.
\bibitem{MAKroupa91} M.A.\ Kroupa, \textit{et al.},
Nucl.\ Inst.\ Meth.\ \textbf{A310}, 649 (1991).
\bibitem{IYLee90} I-Y.\ Lee,
Nucl.\ Phys.\ \textbf{A520}, 641c (1990).
\bibitem{SBaker84} S.\ Baker, \textit{et al.},
Nucl.\ Inst.\ Meth.\ \textbf{221}, 437 (1984).
\bibitem{FJames75} F.\ James, \textit{et al.},
Comp.\ Phys.\ Comm.\ \textbf{10}, 343 (1975).
\bibitem{WRNelson85} W.R.\ Nelson, \textit{et al.},
SLAC Report-265 (1985).
\bibitem{BAWatson69} B.A.\ Watson, \textit{et al.},
Phys.\ Rev.\ \textbf{187}, 1351 (1969).
\bibitem{FAjzenberg-Selove88} F.\ Ajzenberg-Selove,
Nucl.\ Phys.\ \textbf{A490}, 1, (1988).
\bibitem{JFZiegler85} J.F.\ Ziegler, \textit{et al.},
\textit{The Stopping and Range of Ions in Solids},
(Pergamon Press, 1985).
\bibitem{KSiegbahn68} H.\ Frauenfelder and R.M.\ Steffen, in:
$\alpha$-, $\beta$-, and $\gamma$-Ray Spectroscopy,
ed.\ K.\ Siegbahn, (North-Holland, 1968).
\bibitem{DCRobinson72} D.C.\ Robinson, \textit{et al.},
Nucl.\ Phys.\ \textbf{A181}, 645 (1972).
\bibitem{YNagai91} Y.\ Nagai, \textit{et al.},
Phys.\ Rev.\ \textbf{C43}, R9 (1991).
\end{references}
\end{document}